\documentclass[twocolumn,aps,showpacs,groupedaddress]{revtex4}
\usepackage{graphicx}
\usepackage{epsfig}
\usepackage{subfigure}
\usepackage{amsmath}

\begin{document}
%\draft command makes pacs numbers print
%\draft
\title{Nonlinear effect  on quantum control for two-level systems}
\author{W. Wang, J. Shen, and X. X. Yi\footnote{yixx@dlut.edu.cn}}
\affiliation{School of Physics and Optoelectronic Technology,
              Dalian University of Technology, Dalian 116024, China}

\date{\today}

\begin{abstract}
The traditional quantum control theory focuses on linear quantum
system. Here we show the effect of nonlinearity on quantum  control
of a two-level system, we find that the nonlinearity can change the
controllability of  quantum system. Furthermore, we demonstrate that
the Lyapunov control can be used to overcome this uncontrollability
induced by the nonlinear effect.
\end{abstract}

 \maketitle

Quantum control theory is about the application of classical and
modern control strategy  to quantum systems. It has generated
increasing interest in the last few years due to its potential
applications in metrology\cite{wiseman95,berry01},
communications\cite{geremia04a,jacobs07} and other technologies
\cite{ahn02,hopkins03,geremia04b,sarovar04,steck04}, as well as its
theoretical interest. As the effective combination of control theory
and quantum mechanics, the quantum control theory is not trivial for
several reasons. For classical control, feedback is a key factor in
the control design, and there has been a strong emphasis on robust
control of linear  systems. Quantum system in feedback control, on
the other hand, can not usually be modeled as linear control
systems, except when both the system and the controller as well as
their interaction are linear. In fact for many quantum systems, the
nonlinear effects can not be negligible, and in some cases they
dominate the dynamics of quantum system. Moreover, feedback control
requires measurement of an observable and returns the measured
result as a control back to the quantum system. This renders the
dynamics  of the quantum system both nonlinear and
stochastic\cite{habib06}. In special cases the resulting evolution
can be mapped into a linear classical system driven by Gaussian
noise, and consequently the optimal control problem can be solved by
classical control theory. However, most control problems for such a
quantum system can not be solved in this way. Therefore a study on
nonlinear effects in quantum control theory is highly desired.

Lyapunov functions have played a significant role in control design.
Originally used in feedback control to analyze the stability of the
control system, Lyapunov functions  have formed the basis for new
control design. Several papers have be published recently to discuss
the application of  Lyapunov control  to quantum
systems\cite{vettori02,ferrante02,grivopoulos03,
mirrahimi04,mirrahimi05,altafini07,wang08}. Although the basic
mathematical formulism is well established, many questions remain
when one uses the Schr\"odinger equation or the master equation to
describe the dynamics of  quantum system, for example the nonlinear
effect in quantum control.

In this paper, we shall address this issue by analyzing a two-level
system with nonlinear effect. Before studying the nonlinear effect
in quantum control on the two-level system, we recall that a linear
two-level system is controllable by two independent parameters, then
we show that nonlinear interactions may turn the controllable
two-level system into  uncontrollable one. This nonlinear effect may
result from feedback control,  and the uncontrollability can be
overcome by Lyapunov control as we shall show.

Consider a two-level system described by,
\begin{equation}
H=\frac{R}{2}\sigma_{z}+\frac{v}{2}\sigma_{x},\label{h}
\end{equation}
where $\sigma_x$ and $\sigma_z$ are Pauli matrices. This model was
proposed to describe the tunneling of  quantum  system in a
double-well potential. In this model, $v$ is the coupling constant
of the two wells. $R$ denotes the energy difference between the two
levels. We first show that this system is controllable by
manipulating the two independent parameters $R$ and $v$. The
controllability requires all initial states in the Hilbert space
${\cal H}_s$ of the system can evolve to an arbitrary pure target
state. This requirement for the initial state can be partially
lifted by requiring that the Hamiltonian $H$ is unchanged up to $R$
and $v$ under the following unitary transformation\cite{yi08},
\begin{equation}
F=\left(
  \begin{array}{cc}
    \cos\theta & \sin\theta e^{-i\phi} \\
    -\sin\theta e^{i\phi} & \cos\theta \\
  \end{array}
\right),\label{trans}
\end{equation}
where $0\leq \phi \leq 2\pi,$ $0 \leq \theta\leq \pi.$ By unchanged
we mean $H^{\prime}=H(R^{\prime},v^{\prime})=F^{\dagger}H F$, namely
the transformation $F(\theta, \phi)$ changes the control parameter
in the Hamiltonian $H$ only. The proof is straightforward. Consider
the Schr\"odinger equation
$i\hbar\frac{\partial}{\partial{t}}|\varphi(t)\rangle=H|\varphi(t)\rangle,
$ where $|\varphi(t)\rangle$ is the wavefunction of the  system. By
the time-independent transformation $F(\theta,\phi)$ ,
$|\varphi(t)\rangle\rightarrow{F|\phi{'}(t)\rangle}$  we find $
i\hbar\frac{\partial}{\partial{t}}(F|\phi{'}(t)\rangle)=
H(F|\phi{'}(t)\rangle),$ and
$i\hbar\frac{\partial}{\partial{t}}|\phi{'}(t)\rangle=
F^{\dagger}HF|\phi{'}(t)\rangle.$ Since
$H(R^{\prime},v^{\prime})=F^{\dagger}H(R,v)F$, we claim that there
exists a one-to-one correspondence in sets $\{|\phi{'}(t)\rangle\}$
and $\{|\varphi(t)\rangle\}.$ Therefore, if $\{|\phi{'}(t)\rangle\}$
covers all (pure) states in ${\cal H}_s$, $\{|\varphi(t)\rangle\}$
is a convex set of all possible (pure) states for the two-level
system. This observation tells us that if the system initially
prepared in state $|e\rangle$ can be controlled to evolve to an
arbitrary target state driven by the Hamiltonian
$H^{\prime}=H(R^{\prime},v^{\prime})=F^{\dagger}H(R,v)F$, the system
is controllable.
%\vskip -10cm
\begin{figure}
\vskip -2cm
\includegraphics*[width=0.8\columnwidth,
height=1\columnwidth]{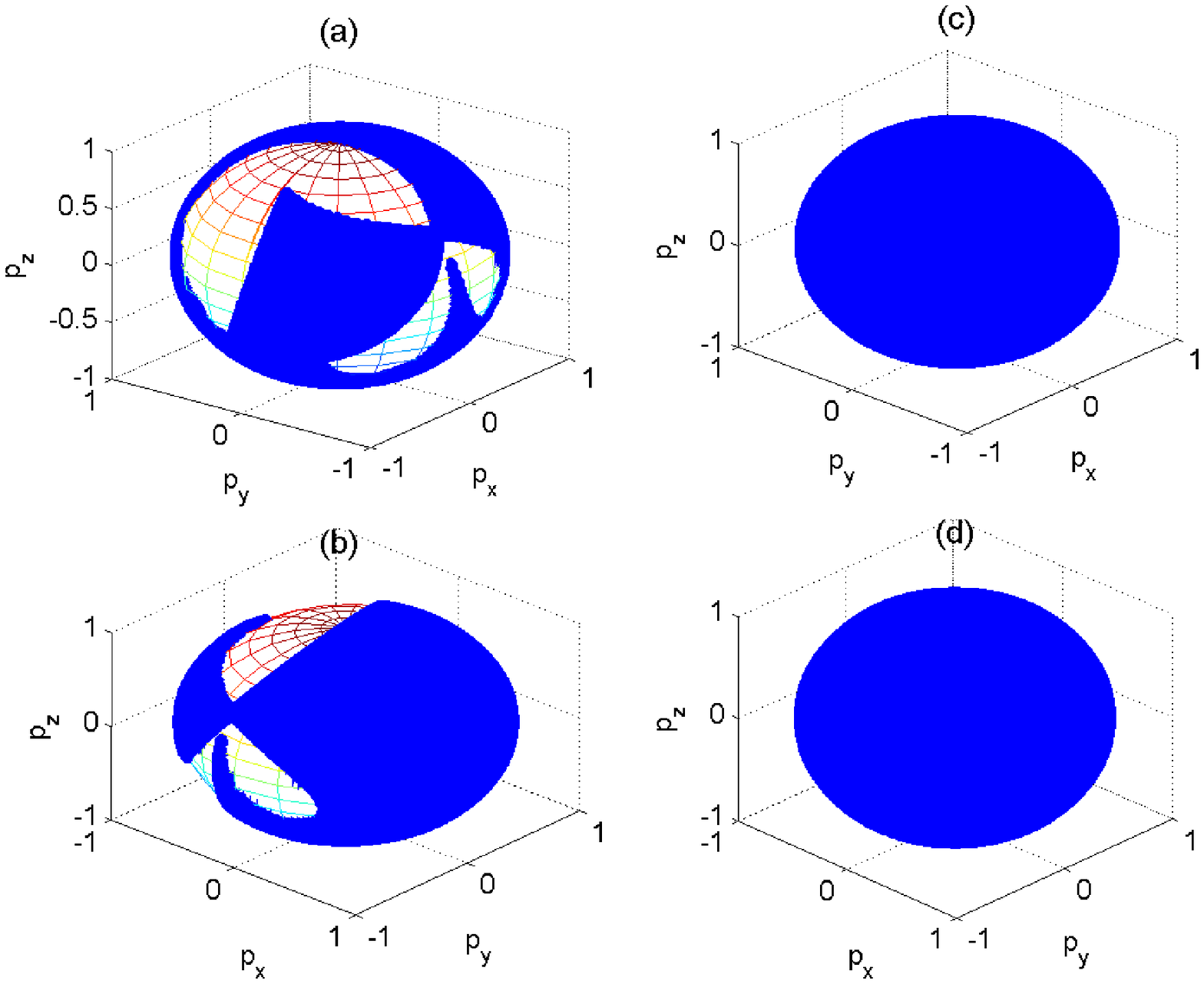} \vskip -2cm \caption{(color
online) Accessible states (blue points on the Bloch sphere) of a
two-level system driven by Hamiltonian Eq.(\ref{h}) in a time
interval $t\in [0,0.5]$ for (a) and (b), and $t\in[0,4]$ for (c) and
(d). The parameters $R$ and $v$ both range from 0 to 7. The initial
state is chosen to be $|e\rangle.$ All parameters are dimensionless.
(a) and (b) [(c) and (d)] are the same but show the results from the
opposite direction. $\vec{p}=(p_x,p_y,p_z)$ is the Bloch vector, and
$\theta=\phi=0.$ } \label{fig1}
\end{figure}
This exactly the case as shown in Fig.\ref{fig1}, where we plot the
accessible states represented by the Bloch vector
$\vec{p}=(p_x,p_y,p_z).$  The Bloch vector is connected to  an
arbitrary state $|\varphi(t)\rangle=a(t)|e\rangle+b(t)|g\rangle$ of
the two-level system through
$\rho=|\varphi(t)\rangle\langle\varphi(t)|=\frac{1}{2}
+\frac{1}{2}\overrightarrow{p}\cdot \overrightarrow{\sigma}$ with
$p_{z}=2|a(t)|^2-1$, $p_{x}=a^*(t)b(t)+b^*(t)a(t),$ and
$p_{y}=i(a(t)b^*(t)-b(t)a^*(t)).$ We find from Fig.\ref{fig1} that
by varying the parameters $R$ and $v$, the two-level system indeed
can evolve to an arbitrary target pure state, provided the evolution
time is long enough and there is a wide range of parameters $R$ and
$v$ to manipulate. It is worth addressing that the Hamiltonian in
Eq.(\ref{h}) becomes
$H^{\prime}=\frac{R^{\prime}}{2}\sigma_z+\frac{v^{\prime}}{2}\sigma_{x}+
\frac{u^{\prime}}{2}\sigma_{y}$ after the unitary transformation
$F(\theta,\phi),$ where $R^{\prime}$ (or $v^{\prime}$) is a function
of $R,v,\theta,$ and $\phi.$ $u^{\prime}$ is in general not zero. At
first sight, $H$ does not satisfy  the condition
$H(R^{\prime},v^{\prime})=F^{\dagger}HF,$ then the two-level system
driven by Hamiltonian Eq.(\ref{h}) is uncontrollable. This is not
the case, however, because there are only two independent parameters
($R$ and $v$) in the Hamiltonian, hence $u^{\prime}$ is not
independent and may be treated as a constant. Therefore the term
with $u^{\prime}$ in $H^{\prime}$ plays no role in the control on
the two-level system\cite{schirmer01,fu01}.

Now we study the effect of nonlinearity on the controllability of
the system. For this goal, we consider a nonlinear two-level model,
\begin{equation}
H_{nl}=\frac{R}{2}\sigma_z-
\frac{C}{2}\langle\psi|\sigma_{z}|\psi\rangle\sigma_{z}+\frac{v}{2}\sigma_{x},
\label{hnl}
\end{equation}
where $|\psi\rangle=|\psi(t)\rangle=\left(
  \begin{array}{c}
    a(t) \\
   b(t)\\
  \end{array}
\right)$ and the parameter $C$ characterizes the nonlinear
interaction strength, and the other parameters have the same
notations as in Eq.(\ref{h}). This model can be used to describe the
tunneling of Bose-Einstein condensates in a double-well potential
and was widely used to study the self-trapping and tunneling in
those systems.

\begin{figure}
\vskip -2cm
\subfigure{\epsfig{file=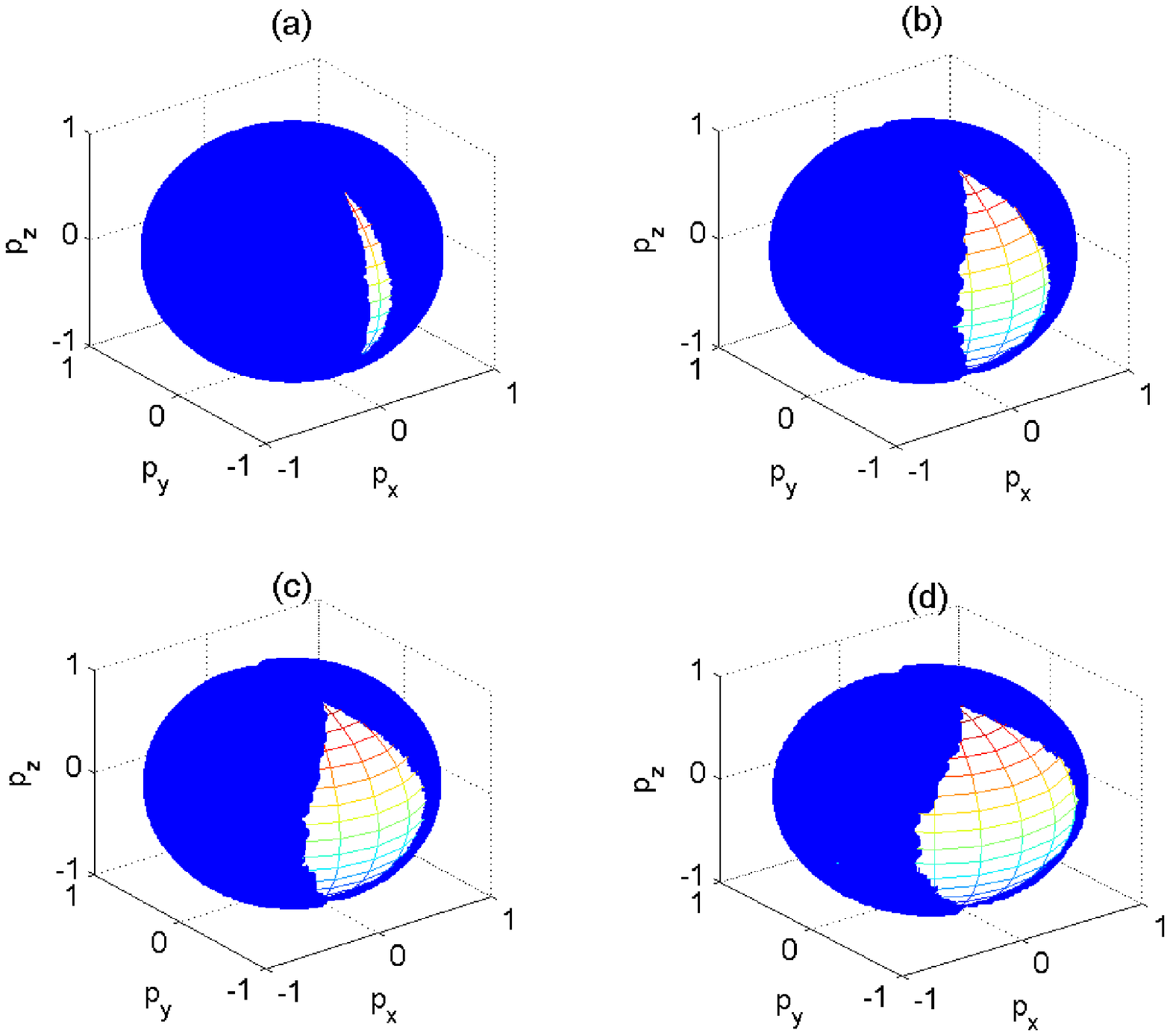,width=0.8\columnwidth,
height=1\columnwidth}} \vskip -5cm
\subfigure{\epsfig{file=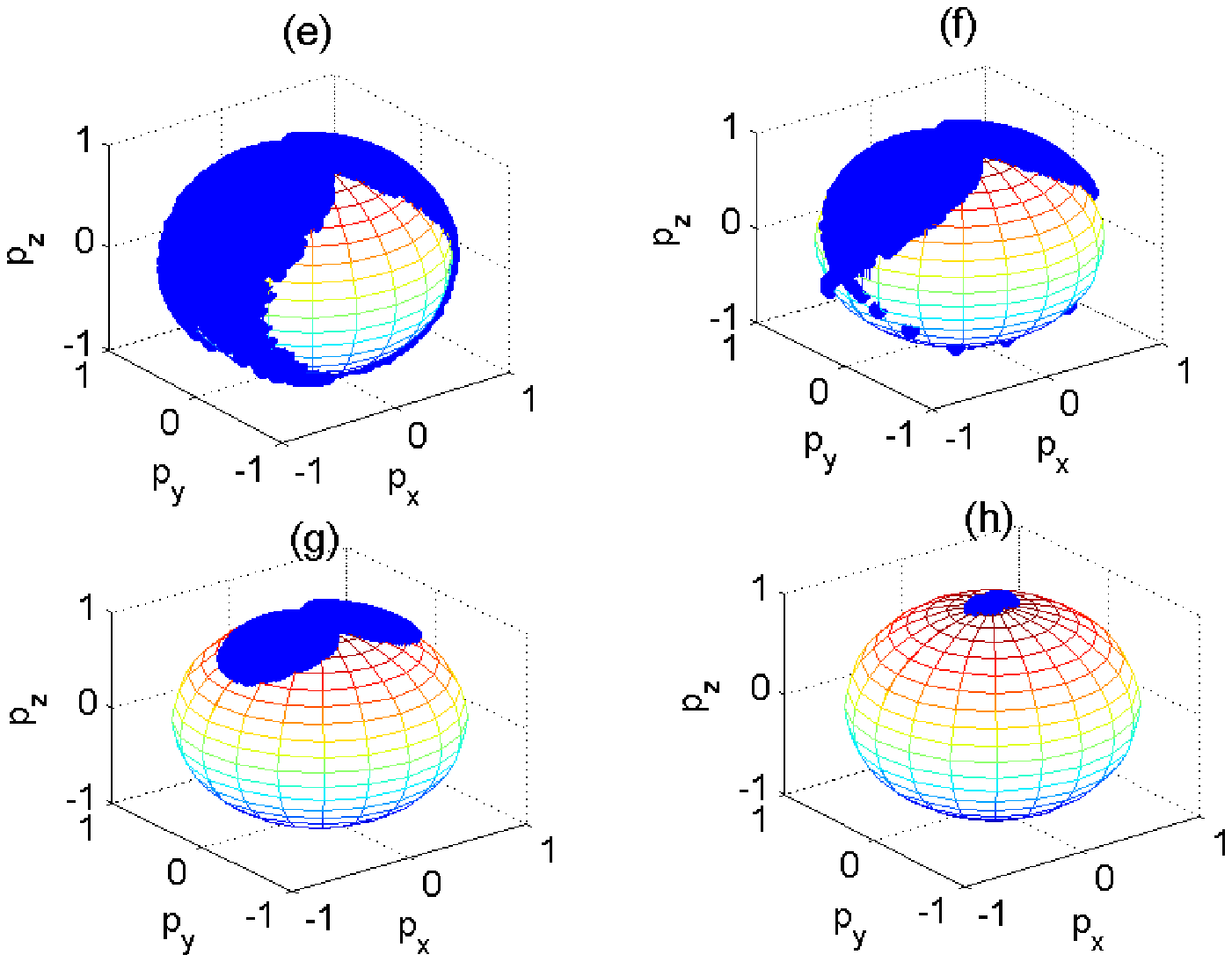,width=1\columnwidth,
height=1.2\columnwidth}} \vskip -3cm \caption{ (color online) Effect
of nonlinearity on the accessible states (blue points on the Bloch
sphere) of a two-level system driven by Eq.(\ref{hnl}) in a time
interval $t\in[0,4].$ Both $R$ and $v$ range from 0 to 7. The axes
$p_x, p_y, p_z$ are the Bloch vectors. The initial state is
$|e\rangle$, i.e., $p_z=1, p_x=p_y=0,$ and $\theta=\phi=0.$ Note
that there is a set of unaccessible states on the other side of the
Bloch sphere, which can not be seen from this angle. (a) $C=2,$ (b)
$C=6,$ (c)$C=8,$ (d) $C=10,$ (e)$C=12,$ (f)$C=14,$ (g)$C=20$ and
(h)$C=100.$} \label{fig2}
\end{figure}

\begin{figure}
\vskip -2cm
\includegraphics*[width=0.7\columnwidth,
height=1\columnwidth]{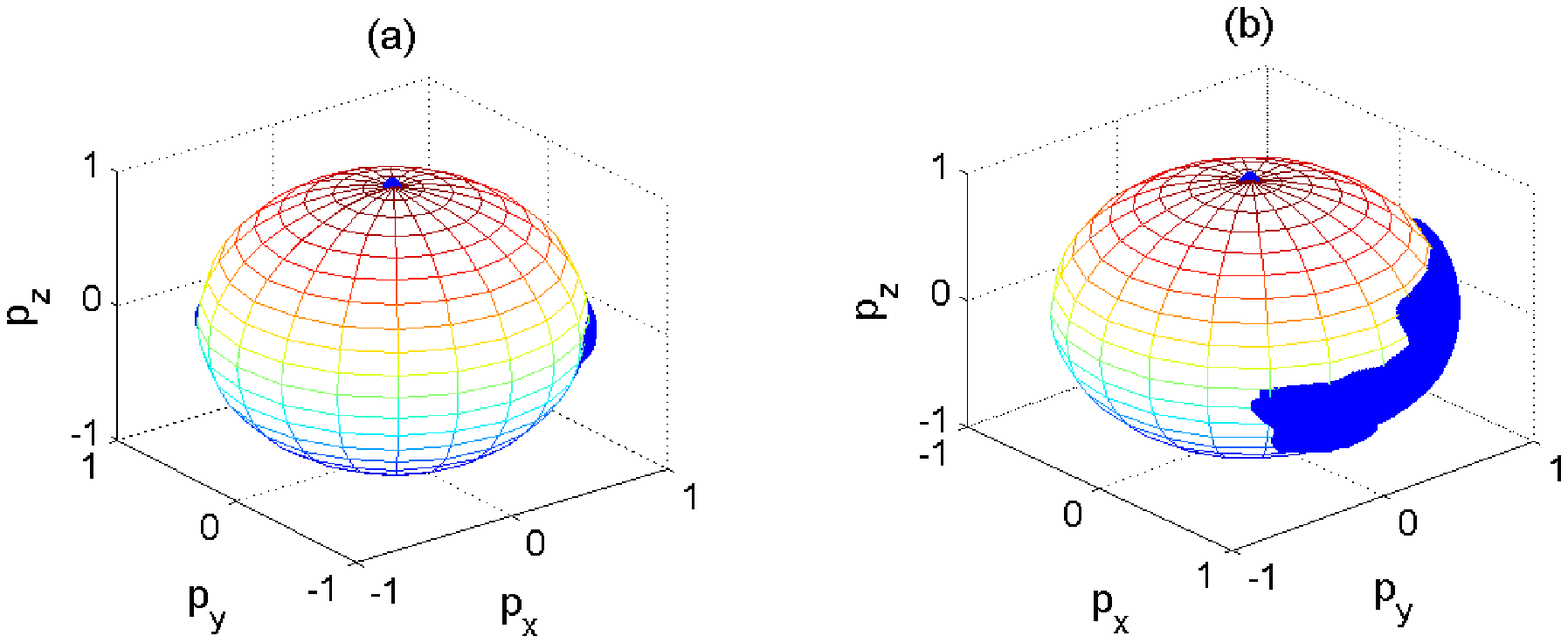} \vskip -3cm \caption{(color
online) The same as  Fig.\ref{fig2}, but with $\theta=\pi/4,
\phi=0.$ $C=20$ was chosen for this plot. (a) is the same as (b),
except the position of axes $p_x$ and $p_y$.} \label{fig3}
\end{figure}

By the unitary transformation $F(\theta,\phi)$, the Hamiltonian
$H_{nl}$ is transformed into,
\begin{eqnarray}
H^{'}_{nl}&=&F^{\dag}H_{nl}F, \\
H^{'}_{nl}&=&(\frac{R_{nl}}{2}\cos^{2}\theta-
v\cos\theta\sin\theta\cos\phi-\frac{R_{nl}}{2}\sin^{2}\theta)\sigma_{z}
\nonumber\\&+&(\frac{v}{2}\cos^{2}\theta+R_{nl}\cos\theta\sin\theta\cos\phi-
\frac{v}{2}\sin^{2}\theta\cos2\phi)\sigma_{x}
\nonumber\\&+&(R_{nl}\cos\theta\sin\theta\sin\phi-
\frac{v}{2}\sin^{2}\theta\sin2\phi)\sigma_{y},\\
\end{eqnarray}
where $R_{nl}=R-C\langle\psi|\sigma_{z}|\psi\rangle.$ We have
performed extensive numerical simulations for the Schr\"odinger
equation $i\hbar\frac{\partial}{\partial
t}|\psi\rangle=H^{\prime}_{nl}|\psi\rangle$ with $\theta=\phi=0,$
select results are presented in Fig.\ref{fig2}.
Two observations can be made from Fig. \ref{fig2}. (1) The
nonlinearity affects the controllability of the two-level system,
regardless of how small the nonlinear coupling constant $C$ is, (2)
the larger the nonlinear term is, the smaller the set of the
accessible states. Further numerical simulation shows that $\theta$
and $\phi$ (determining the initial state) can change the accessible
set of states but not the two observations, this is shown in
Fig.\ref{fig3}.

The nonlinearity in  nonlinear quantum system may come from feedback
control $\frac{C}{2}\langle\psi|\sigma_{z}|\psi\rangle\sigma_{z}$.
This feedback control may be understood as applying a pulse to shift
the energy level of the quantum system, the amplitude of the pulse
is proportional to a detection event with a rate $C$ described by
$\langle \psi|\sigma_z|\psi\rangle$.
 To overcome the uncontrollability induced by the feedback, we
introduce a Lyapunov control $f(t)$ to replace the coupling constant
$C$ in the nonlinear term
$\frac{C}{2}\langle\psi|\sigma_{z}|\psi\rangle\sigma_{z}.$ The
Lyapunov control may be designed according to the Lyapunov control
theory as follows\cite{wang08}. For a two-level system, its
Hamiltonian can be rewritten as,
\begin{eqnarray}
H_{nl}=H_{0}+f(t)H_{1}, \label{hly}
\end{eqnarray}
where $H_0$ is the free evolution Hamiltonian and $H_1$ is the
control Hamiltonian. The general control task we consider can be
formulated as, given a target state $|\psi_d(t)\rangle$, we wish to
apply a certain control field $f(t)$ to the system that modifies its
dynamics such that $|\psi(t)\rangle\rightarrow |\psi_d(t)\rangle$ as
$t\rightarrow \infty.$ Since the free Hamiltonian can in general not
be turned off, it is natural to assume $|\psi_d(t)\rangle$ to be
time-dependent and satisfies
\begin{equation}
i \hbar\frac{\partial}{\partial
t}|\psi_d(t)\rangle=H_0|\psi_d(t)\rangle.
\end{equation}
Since the evolution of both $|\psi(t)\rangle$ and
$|\psi_d(t)\rangle$ are unitary in our case, we can define a
function
\begin{equation}
V[|\psi_d(t)\rangle, |\psi(t)\rangle]=1-|\langle
\psi_d(t)|\psi(t)\rangle|^2
\end{equation}
to measure the distance between the resulting and target states.
Clearly $V \geq 0$ with equality only if
$|\psi_d(t)\rangle=|\psi(t)\rangle$. Taking derivative of $V$ with
respect to time $t$, we have ($|\psi_d\rangle=|\psi_d(t)\rangle$ and
$|\psi\rangle=|\psi(t)\rangle$ hereafter)
\begin{equation}
\dot{V}=-2f(t)\mbox
{Im}(\langle\psi_d|H_1|\psi\rangle\langle\psi|\psi_d\rangle),
\end{equation}
where $\mbox{Im}(...)$ denote the imaginary part of $(...).$   So
when we choose $f(t)=\kappa \mbox
{Im}(\langle\psi_d|H_1|\psi\rangle\langle\psi|\psi_d\rangle)$ with a
rate $\kappa>0$, we have $\dot{V}\leq 0.$ Therefore $V$ is a
Lyapunov function for the following dynamical system,
\begin{eqnarray}
i\hbar\frac{\partial}{\partial
t}|\psi\rangle&=&H_{nl}|\psi\rangle,\nonumber\\
i\hbar\frac{\partial}{\partial
t}|\psi_d\rangle&=&H_0|\psi_d\rangle,\nonumber\\
f(t)&=&\kappa \mbox
{Im}(\langle\psi_d|H_1|\psi\rangle\langle\psi|\psi_d\rangle).
\end{eqnarray}
For a two-level system, $|\psi_d\rangle$ always can be written as
$|\psi_{d}\rangle=c(t)|e\rangle+d(t)|g\rangle$ with
$|c(t)|^2+|d(t)|^2=1$, and
$|\psi\rangle=a(t)|e\rangle+b(t)|g\rangle$ with
$|a(t)|^2+|b(t)|^2=1.$ When the control Hamiltonian takes
$H_{1}=\sigma_{z}/2$, it is easy to find that the Lyapunov control
$f(t)=-2\kappa \mbox{Im}(cd^*a^*b).$ In the following, we shall
focus on the control Hamiltonian
$H_{1}=\frac{1}{2}\langle\psi|\sigma_{z}|\psi\rangle\sigma_{z},$
which yields the Lyapunov control
\begin{eqnarray}
f(t)&=&-\kappa m \mbox{Im}[(ac^{*}+bd^*)(a^*c-b^*d)],\nonumber\\
m&=&|a|^2-|b|^2,
\end{eqnarray}
where we omitted the argument $t$ of $a(t), b(t), c(t)$ and $d(t)$
to shorten the notations. Clearly, the Lyapunov control renders the
dynamics of the quantum system nonlinear even if the control
Hamiltonian is linear. We have performed extensive numerical
simulations for the dynamics of these nonlinear system, the
numerical simulations show that the two-level system described by
Eq.(\ref{h}) with Lyapunov control  $f(t)$ is controllable, namely
an arbitrary pure state is accessible driven by
\begin{equation}
H_{nl}=\frac{R}{2}\sigma_z+\frac{v}{2}\sigma_{x}-
\frac{f(t)}{2}\langle\psi|\sigma_{z}|\psi\rangle\sigma_{z}.
\label{hly}
\end{equation}
Then a natural question arises  for $\kappa,$ how does the rate
$\kappa$ in Lyapunov control $f(t)$ affect the accessible  set of
states?
\begin{figure}
\vskip -2cm
\includegraphics*[width=0.9\columnwidth,
height=1.2\columnwidth]{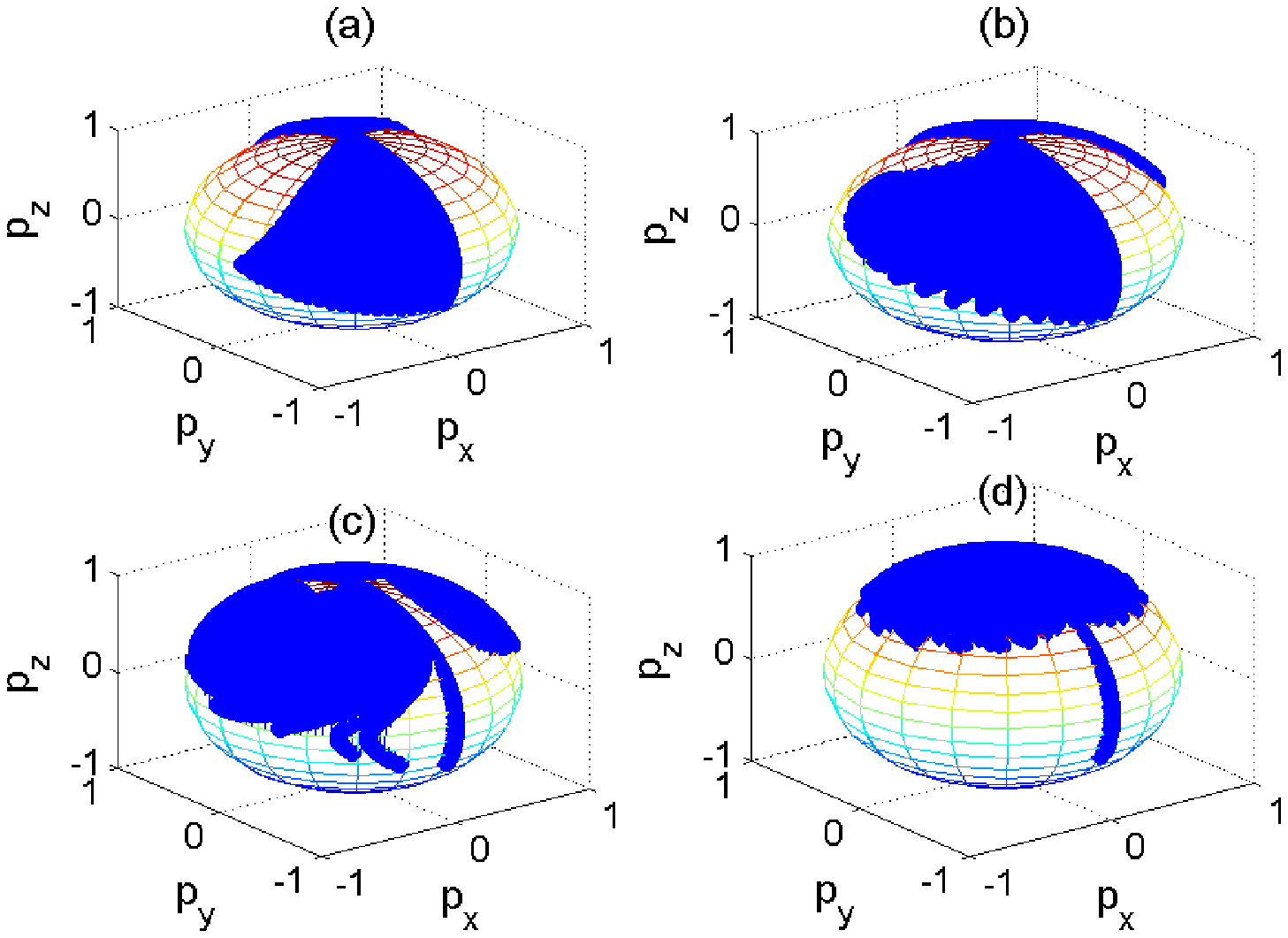} \vskip -3cm
\caption{(color online) Accessible states represented by the Bloch
vector $\vec{p}=(p_x,p_y,p_z)$. The two-level system is driven by
Eq.(\ref{hly}) with Lyapunov feedback control $f(t)$ evolving in a
time interval $t\in[0,4]$ with $\theta=\phi=0$. $R$ and $v$ range
from 0 to 0.5. (a) $\kappa=0$, (b) $\kappa=3$, (c) $\kappa=9$, (d)
$\kappa=27$. } \label{fig4}
\end{figure}
Fig. \ref{fig4} shows the sets of accessible state reached by
controlling the parameter $R$ and $v$ in a small range. In other
words, the control parameters $R$ and $v$ are restricted in a regime
much smaller than that by which the two-level system is
controllable. We find that the rate $\kappa$ affects the accessible
states. For $\kappa$ below a critical value $\kappa_c$, the larger
the $\kappa$ is, the bigger the set of accessible states. For
$\kappa>\kappa_c,$ the situation changes, smaller $\kappa$ favors
the set of accessible states.  In Fig.\ref{fig5}, we plot the
short-time behavior of the accessible states, observations  similar
to Fig.\ref{fig4} can be found. A common feature we find from
Fig.\ref{fig4} and \ref{fig5} is that as $\kappa$ increases, states
near the initial state $|e\rangle$ become easy to access. This
finding depends on $\theta$ and $\phi.$ Finally, we address the
convergence for the control system. By the Lasalle's invariance
principle\cite{lasalle61}, the largest invariant set is empty, so
there is not any invariant set for the problem under consideration.
The reason is that we choose $|\psi_d(t=0)\rangle=|\psi(t=0)\rangle$
throughout this paper.
\begin{figure}
\vskip -2cm
\includegraphics*[width=0.8\columnwidth,
height=1\columnwidth]{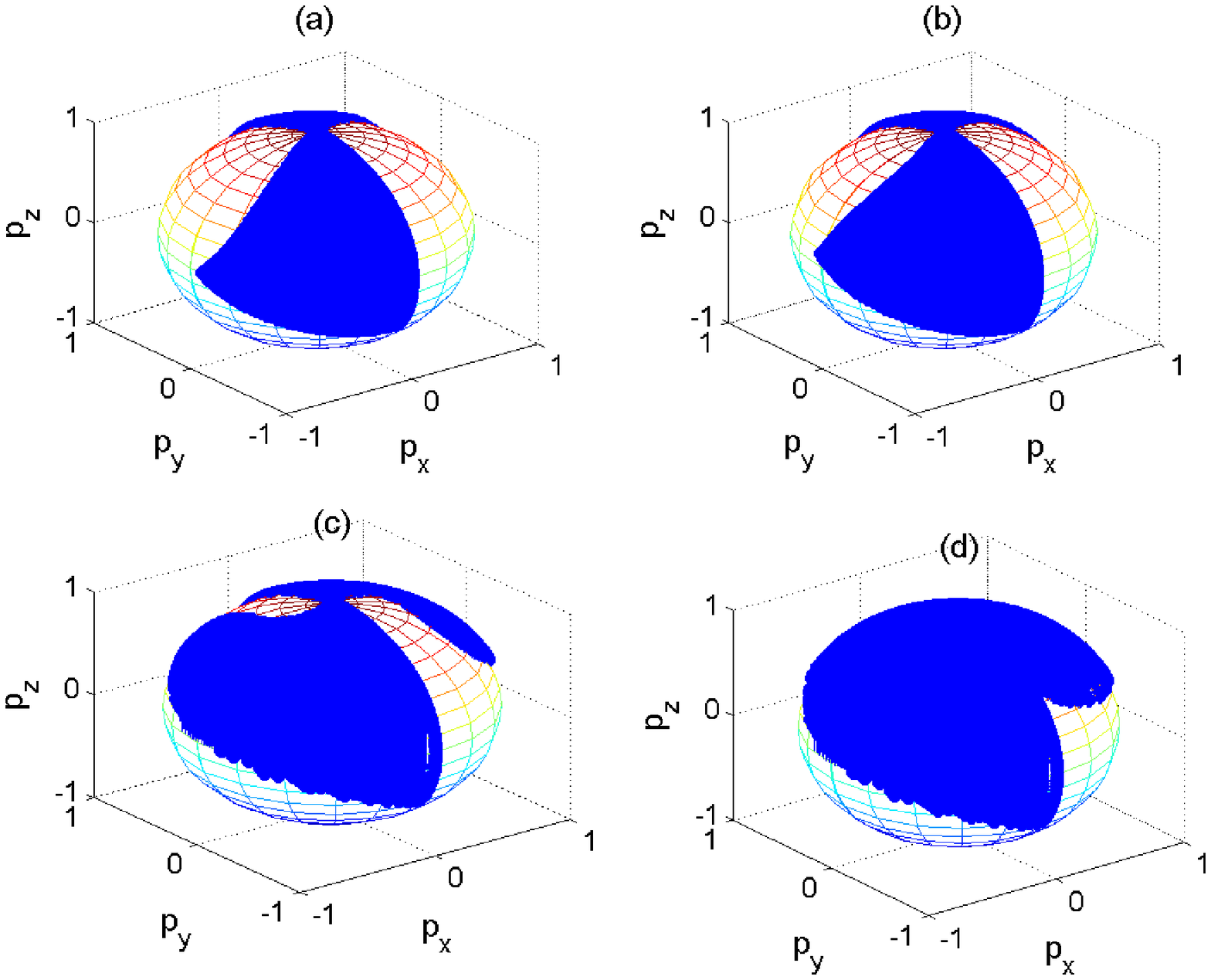} \vskip -2cm \caption{(color
online) The same as  Fig.\ref{fig4}, but the time interval is
$[0,0.3]$, and $R$ and $v$ vary  in $[0,7]$. (a) $\kappa=0$, (b)
$\kappa=9$, (c) $\kappa=81$, (d) $\kappa=243$.} \label{fig5}
\end{figure}

In summary, we have investigated the nonlinear effect on the
controllability of a two-level system. This nonlinear effect can
turn a controllable quantum system uncontrollable. The accessible
set of states under nonlinear effect depends on the nonlinear
coefficient $C$ and the initial state. To overcome this
uncontrollability induced by the nonlinear effect, we propose to use
Lyapunov control to manipulate the two-level system, Lyapunov
function for the control system is constructed and the dependence of
accessible set of states, which can be reached in a short-time limit
and within a small range of control parameters on the rate $\kappa$
are shown and discussed. This study suggests that feedback control
that can induce nonlinear effect  changes the controllability of
quantum system, Lyapunov control is better in this case for
manipulating a quantum system.

This work was supported by   NSF of China under Grant  No. 10775023.\\

\end{document}